\documentclass[
preprint,
 amsmath,amssymb,
 aps,prl,
]{revtex4-1}
\usepackage{float} 
\usepackage{color}
\usepackage{fullpage}
\usepackage{graphicx}
\usepackage{dcolumn}
\usepackage{bm}
\usepackage{hyperref}

\begin{document}

\title{Characterizing Hydration Properties Based on the Orientational Structure of Interfacial Water Molecules}

\author{Sucheol Shin}
\author{Adam P. Willard}
\email{awillard@mit.edu}
\affiliation{ 
Department of Chemistry, Massachusetts Institute of Technology, Cambridge, Massachusetts 02139, USA
}

\begin{abstract}
In this manuscript, we present a general computational method for characterizing the molecular structure of liquid water interfaces as sampled from atomistic simulations. With this method, the interfacial structure is quantified based on the statistical analysis of the orientational configurations of interfacial water molecules. The method can be applied to generate position dependent maps of the hydration properties of heterogeneous surfaces. We present an application to the characterization of surface hydrophobicity, which we use to analyze simulations of a hydrated protein. We demonstrate that this approach is capable of revealing microscopic details of the collective dynamics of a protein hydration shell.
\end{abstract}

\maketitle

In the vicinity of a hydrated surface the properties of water can differ significantly from that of the bulk liquid \cite{Eisenthal:1996jh}. These differences are determined by the details of the interfacial environment and they are thus sensitive to the specific chemical and topological features of the hydrated surface \cite{Quere:2008cw}. The details of water's interfacial molecular structure, therefore, contain information about these surface features and how they affect their local aqueous environment \cite{Verdaguer:2006gl}. This information is valuable because it provides insight into the collective molecular effects that control the solvation properties of complex solutes, but it is also difficult to access due to limitations in our ability to measure and characterize water's interfacial molecular structure. 

Our current understanding of the interfacial molecular structure of liquid water is derived primarily from the results of interface sensitive experimental techniques such as vibrational sum frequency generation spectroscopy \cite{Du:1993je, Schleeger:2014ek, Singh:2012kc, Baldelli:2002be, Gragson:1997dj}, THz absorption spectroscopy \cite{ContiNibali:2014gt, Ebbinghaus:2007gl}, dynamic nuclear polarization \cite{Armstrong:2009et}, NMR \cite{Otting:1989jv, Otting:1991ue}, and X-ray and neutron scattering \cite{Svergun:1998ej, Bruni:1998gy, Luo:2006gu}. Unfortunately, these experiments are typically more difficult to interpret than their bulk phase counterparts due to the constraints associated with achieving interface selectivity. This has driven an increased demand for theoretical developments that can facilitate the analysis and interpretation of these interface-sensitive experiments. Resulting efforts have relied heavily on the use of atomistic simulation to provide the molecular details of water's interfacial structure.

Classical molecular dynamics (MD) simulations are a particularly efficient theoretical framework for modeling the microscopic properties of aqueous interfacial systems. These simulations provide a molecular-level basis for understanding the microscopic structure of liquid water and how it responds to the anisotropic environment of the liquid phase boundary. This response is mediated by the properties of water's hydrogen bonding network and therefore involves the correlated arrangements of many individual water molecules. Characterizing this high-dimensional molecular structure in simple and intuitive terms can be a significant challenge, especially for solutes such as proteins that exhibit irregular or heterogeneous surface properties. 

Here we address this challenge by characterizing water's interfacial molecular structure in terms of a low dimensional parameter that quantified its similarity to the structure of various interfacial reference systems. We introduce a theoretical framework for quantifying this similarity based on the statistical analysis of molecular orientations at the interface. In this analysis reference systems serve to designate the unique orientational signatures of water interfaces at surfaces with specific well-defined chemical or topographical properties. The properties of these reference surfaces can be systematically chosen in order to analyze specific interfacial features that may be relevant to a particular system of study. This framework provides a physically insightful measure of interfacial structure that eliminates the need to formulate high-dimensional collective variables. Furthermore, by applying this framework across a variety of different reference systems, it can be adapted to report simultaneously on multiple specific interfacial properties. 

The general formalism for our framework begins with the definition of the reference system(s) that will serve as a basis for interfacial characterization. Prior to applying this framework each reference system must be thoroughly sampled in order to establish its unique orientational molecular signature. We quantify this signature in terms of the molecular orientational distribution function, 
\begin{equation}
f(\vec{\kappa}|\text{ref}) = \frac{P(\vec{\kappa}|\text{ref})}{P(\vec{\kappa}|\text{bulk})}~,
\end{equation}
where $P(\vec{\kappa}|\text{ref})$ and $P(\vec{\kappa}|\text{bulk})$ denote the equilibrium probabilities for observing a molecule with a specific molecular configuration, $\vec{\kappa}$, within the given reference system ands in the bulk liquid respectively. We specify molecular configuration in terms of a three-dimensional vector, $\vec{\kappa}=(\cos\theta_1, \cos\theta_2, a)$, where $\theta_1$ and $\theta_2$ denote the angles made between the local surface normal and each of a water molecule's OH bond vectors, and a denotes the distance of the water molecule from the instantaneous position of the water interface. We define the instantaneous water interface following the procedure of Ref. \cite{Willard:2010da}. This system of molecular coordinates is illustrated schematically in Fig. 1.

\begin{figure}[t!]
\centering
\includegraphics[width = 3.4 in]{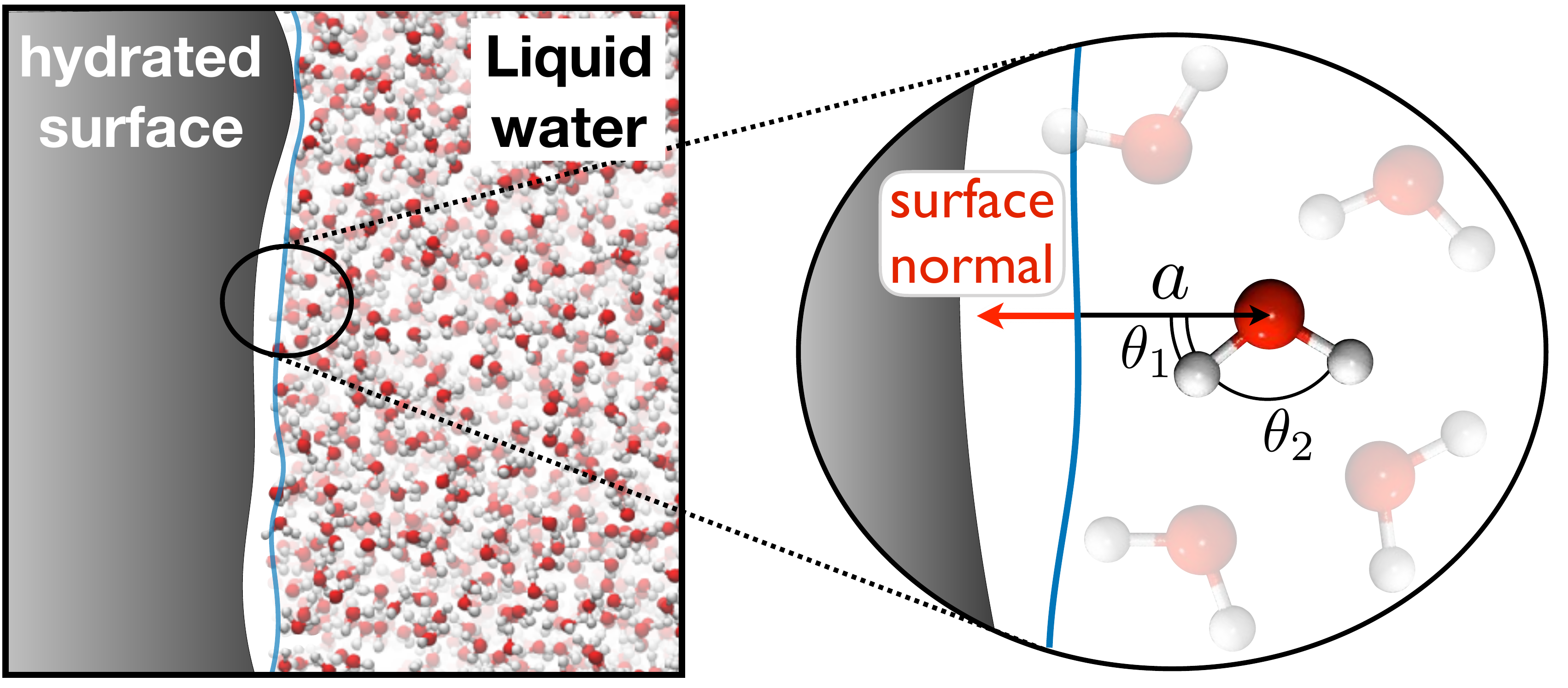}
\caption{A schematic illustration of the molecular coordinates that are used to specify the orientational state, $\vec{\kappa}$, of molecules at the liquid water interface.}
\label{fig:1}
\end{figure}

We characterize the interfacial molecular structure of of a particular interfacial system by sampling values of $\vec{\kappa}$ and comparing them to the distribution function, $f(\vec{\kappa}|\text{ref})$. Specifically, we compute the quantity, 
\begin{equation}
\lambda_\text{ref} = \frac{1}{N}\sum_{i=1}^{N}- \ln\left[f(\vec{\kappa}|\text{ref})\right]~,
\label{eq:2}
\end{equation}
where the summation is taken over a set of $N$ molecular configurations sampled from the system of interest. This quantity reflects the likelihood for the sampled set of configurations to occur spontaneously within the environment of the reference system. This likelihood is relatively large (corresponding to lower values of $\lambda_\text{ref}$) when interfacial molecular structure is similar to that of the reference system and relatively small (corresponding to higher values of $\lambda_\text{ref}$) when interfacial molecular structure differs from that of the reference system. In this way, $\lambda_\text{ref}$ can be used to distinguish hydrated surfaces based on how they influence their hydration environment, irrespective of their specific chemical or topographical properties.  

For heterogeneous surfaces, this framework can be applied locally to generate spatially resolved maps of $\lambda_\text{ref}$. This can be accomplished by restricting sampling of $\vec{\kappa}$ to specific systematically controllable regions along the surface. A general expression for Eq. (\ref{eq:2}) that can be used to compute $\lambda_\text{ref}$ at a specific position, $\vec{r}_\text{surf}$, and time, $t$ is given by,
\begin{equation}
\lambda_\text{ref}(\vec{r}_\text{surf},t) = \frac{1}{\tau}\sum_{t^\prime=t}^{t+\tau}- \ln\left[f(\vec{\kappa}(\vec{r}_\text{surf},t^\prime)|\text{ref})\right]~,
\label{eq:3}
\end{equation}
here $\vec{\kappa}(\vec{r}_\text{surf},t^\prime)$ is the orientational configuration of the water molecule that is closest to position $\vec{r}_\text{surf}$ at time $t^\prime$, and the summation is taken over a series of $\tau$ discrete time steps. The value of $\tau$ in Eq. (\ref{eq:3}) can be varied to highlight average interfacial response to a heterogeneous surface (\emph{i.e.}, large $\tau$) or to highlight transient fluctuations in interfacial molecular structure (\emph{i.e.}, small $\tau$). 

Early approaches to mapping the hydration properties of heterogeneous solutes, most notably those developed by Kyte and Doolittle, were based on a spatial decomposition of surface chemistry \cite{Kyte:1982fr}, and have since been extended to provide higher resolution \cite{Kapcha:2014fg}. These approaches often fail to accurately predict solvation properties due to their neglect of transverse correlations within the water interface. More recent approaches have focused on water-based mapping methods. This includes approaches based on local density fluctuations \cite{Limmer:2015bq, Patel:2011dz}, single water chemical potentials \cite{Young:2007cx}, and local electrostatic fields \cite{Remsing:2015dh}. Our method is complementary to these previous approaches and can be adapted, via the selection of different reference systems, to map a wide variety of interfacial properties.

We illustrate the performance of our method by applying it to characterize the hydrophobicity of heterogeneous surfaces. To do this we use a single reference system as a basis for interfacial characterization -- the liquid water interface at an ideal hydrophobic surface. The collective molecular arrangements that are common to this reference system are thus specified by $f(\vec{\kappa}|\text{phob})$. As we demonstrate below, and in the Supporting Information (SI), $\lambda_\text{phob}$ is capable of distinguishing between the interfacial molecular structure of hydrophobic and hydrophilic surfaces and can thus be treated as an order parameter for hydrophobicity. As such, we use $\lambda_\text{phob}$, computed according to Eq. (\ref{eq:3}), to analyze water's interfacial response to heterogeneous surfaces. Since $\lambda_\text{phob}$  is based only on water's interfacial molecular structure, it can be used to generate hydrophobicity maps that reveal the effective solvation characteristics of surfaces with complex or unknown properties.

As a proof of concept, we apply our framework for interfacial characterization to a model silica surface with a patterned composition of hydrophobic and hydrophilic surface sites \cite{Giovambattista:2007cj}. As illustrated in Fig. 2, the surface sites of this model can be either nonpolar (\emph{i.e.}, hydrophobic), if they are terminated with a neutral silica atom, or polar (\emph{i.e.}, hydrophilic), if they are terminated with a charged hydroxyl group. Artificial surfaces with well-defined surface patterns can be created by specifying the hydroxylation state of the surface sites. We then use $\lambda_\text{phob}$ to analyze water's response to various surface patterns. We quantify this response by computing $\delta \lambda_\text{ref} = \lambda_\text{ref} - \langle \lambda_\text{ref} \rangle_0$, where $\langle \cdots \rangle_0$ denotes an equilibrium average taken within the ensemble of configurations sampled directly from the reference system. In this way, values of $\delta\lambda_\text{ref} \approx 0$ correspond to interfacial environments that resemble that of the reference system. Further details about the simulations are described within the SI.

\begin{figure}[t!]
\centering
\includegraphics[width = 3.4 in]{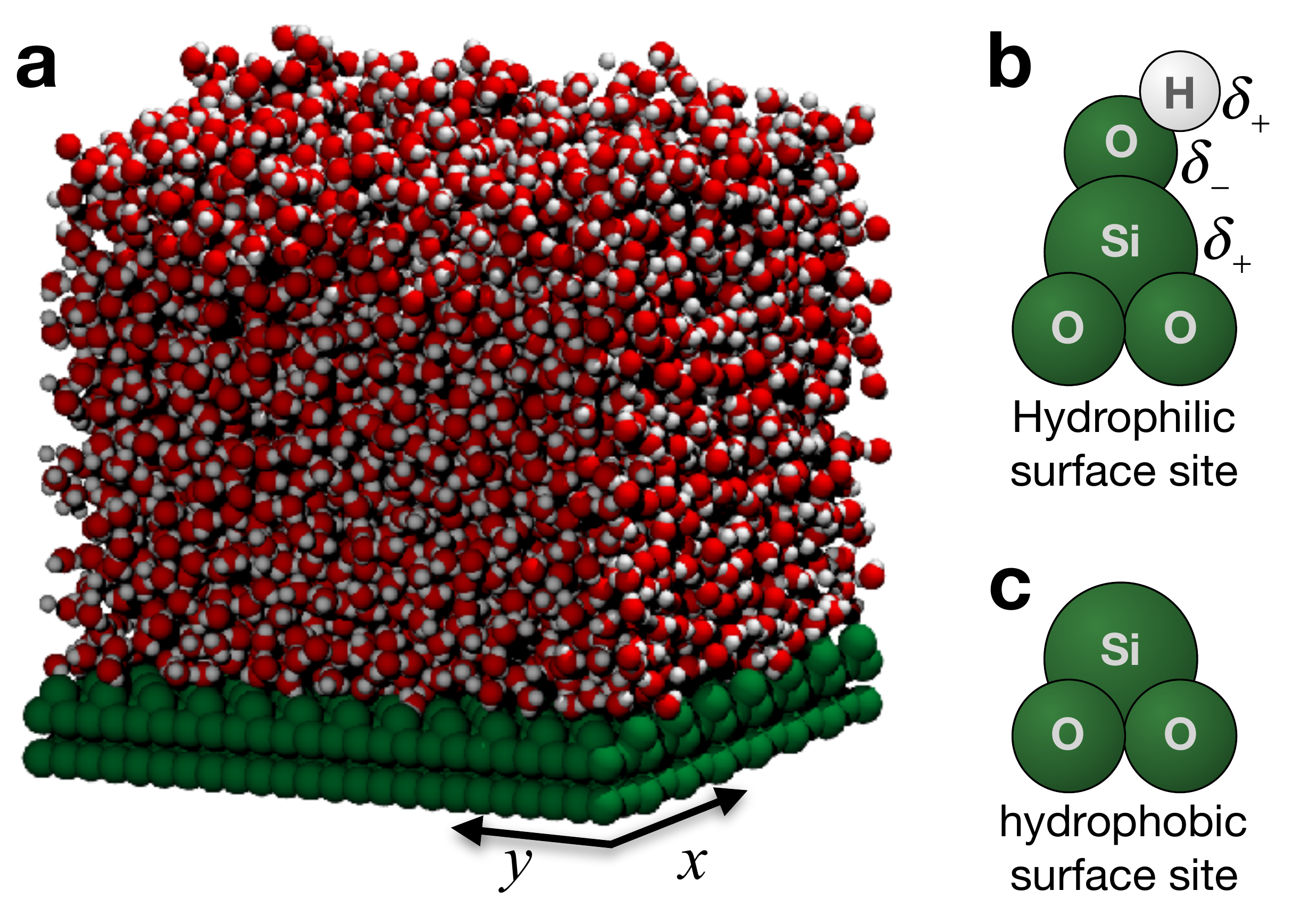}
\caption{(a) A snapshot of a simulation of a periodically replicated slab of liquid water in contact with a $6 \times 6 \text{ nm}^2$ model silica surface. (b) and (c) The chemical termination of the surface sites determines whether they are hydrophobic or hydrophilic.}
\label{fig:2}
\end{figure}

Figure 3 highlights the ability of $\lambda_\text{phob}$ to distinguish between water's interfacial molecular response to hydrophobic and hydrophilic regions of a spatially heterogeneous surface. Specifically, we have computed $\lambda_\text{phob}$, using Eq. (\ref{eq:3}), for points along an extended hydrophobic surface with a larger rectangular hydrophilic patch, as shown in Fig. 3a. For the data plotted in Fig. 3b, $\lambda_\text{phob}$ has been averaged over a long observation time of $\tau = 4 \text{ ns}$ (40000 individual configurations). We observe that water's average interfacial molecular structure exhibits spatial variations that mimic the chemical patterning of the underlying silica surface. Over non-polar regions of the surface $\delta\lambda_\text{phob} \approx 0$ (\emph{i.e.}, $\lambda_\text{phob} \approx \langle \lambda_\text{phob} \rangle_0$), indicating that interfacial molecular structure is similar to that of the hydrophobic reference system.  Over polar regions of the surface $\delta\lambda_\text{phob} > 0$ (\emph{i.e.}, $\lambda_\text{phob} > \langle \lambda_\text{phob} \rangle_0$), indicating that water's interfacial response differs significantly from that of the reference system. Low amplitude spatial modulations in $\lambda_\text{phob}$ can be observed over both the polar and the non-polar regions of the surface. These modulations reflect the corrugation of the atomic surface and thus indicate that this order parameter is sensitive to the subtle influence of surface topography on water's interfacial molecular structure.

Transient fluctuations in local interfacial structure can be analyzed by computing $\delta\lambda_\text{phob}$ with a smaller value of $\tau$. For instance, Fig. 3d shows $\delta\lambda_\text{phob}$ computed for the surface in Fig. 3a using a value of $\tau = 20 \text{ ps}$ (100 individual configurations). The use of a shorter observation time highlights the transient details of water's interfacial molecular structure. Thus, by comparing local values of $\delta\lambda_\text{phob}$ over multiple consecutive snapshots it is possible to observe the transient fluctuations of interfacial molecular structure and investigate how they depend on the details of local surface chemistry and topology. The dynamic range of $\delta\lambda_\text{phob}$ within the hydrophobic reference system is described in the SI.

\begin{figure}[t!]
\centering
\includegraphics[width = 3.4 in]{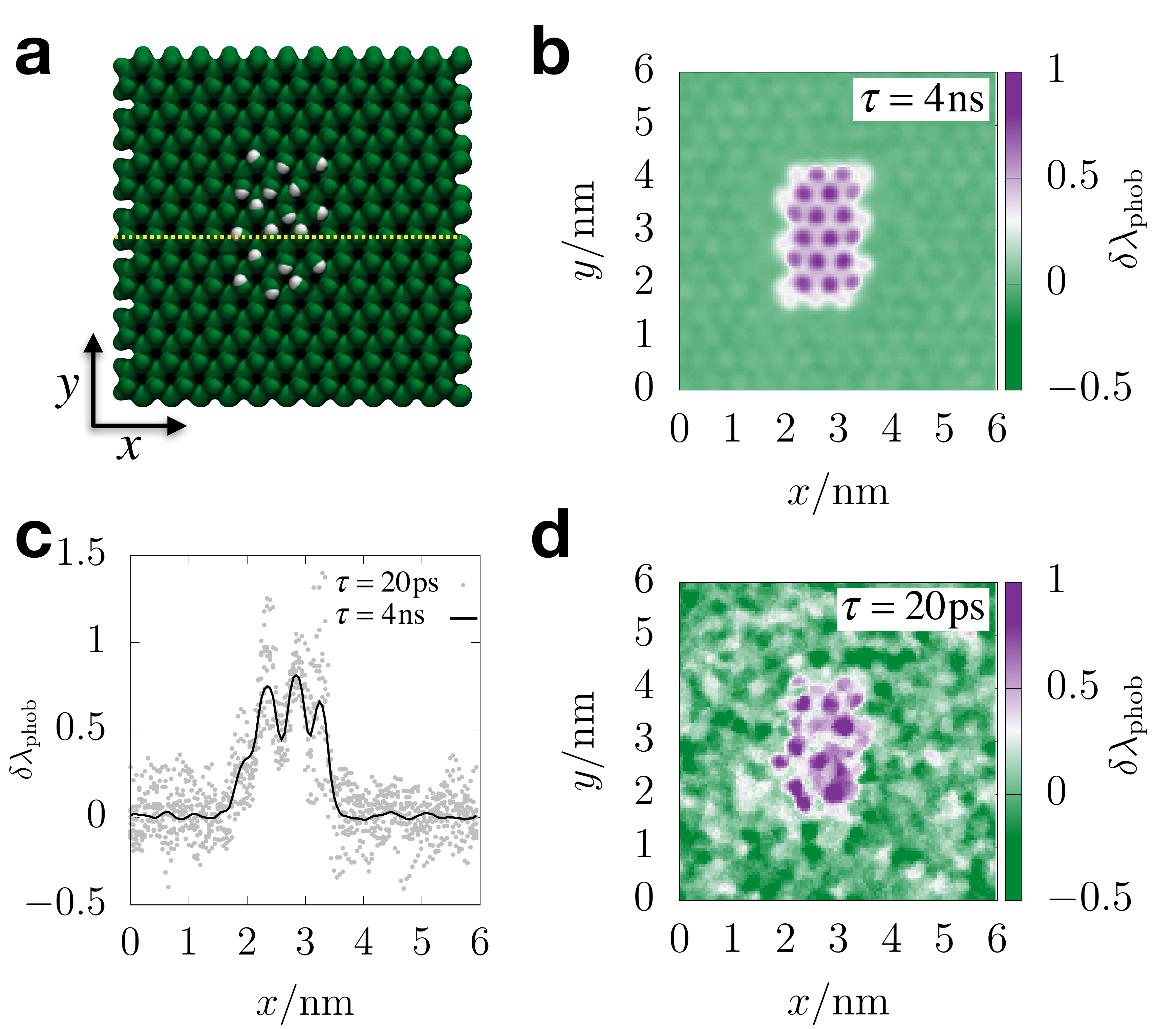}
\caption{(a) A snapshot of the water exposed face of a model silica surface featuring a hydrophilic patch against a hydrophobic background. (b) and (d) A plot of $\delta\lambda_\text{phob}$, indicated by shading, computed for points along the surface of the structure shown in Panel (a). The data in Panels (b) and (d) reflects an average over an observation time of $\tau = 4 \text{ ns}$ and $\tau = 20 \text{ ps}$, respectively. (c) A plot of the value of $\delta\lambda_\text{phob}$ computed along a line at $y = 3 \text{ nm}$ (the yellow dotted line in Panel (a)) that highlights how interfacial molecular structure is affected by the patch boundary. Grey points are values of $\delta\lambda_\text{phob}$, computed with $\tau = 20 \text{ ps}$, sampled at different points in time and the solid line is $\delta\lambda_\text{phob}$ computed with $\tau = 4 \text{ ns}$.}
\label{fig:3}
\end{figure}

Water molecules that reside over the boundaries between polar and non-polar regions of the surface experience a laterally anisotropic aqueous environment. In these regions, such as along the edge of the polar patch of the surface illustrated in Fig. 3a, $\delta\lambda_\text{phob}$ takes on values that are intermediate between that of the extended polar and non-polar surfaces. The characteristics of $\delta\lambda_\text{phob}$ in these boundary regions reveal details about the molecular correlations that mediate interactions along liquid water interfaces, and how these correlations are influenced by the details of surface-water interactions. 

In the case of the model silica surface we observe that the effect of the polar/non-polar surface boundary on the interfacial molecular structure is local, limited to the region directly above the surface boundary. A cross-section of $\delta\lambda_\text{phob}$ that cuts through the center of the hydrophilic surface patch is plotted in Fig. 3c. This plot reveals that the influence of a large hydrophilic surface patch on water's interfacial molecular structure only extends about one molecular diameter beyond the edge of the patch. Evidently, in this case any long-ranged transverse correlations due to heterogeneous surface chemistry are not supported by distortions of the interfacial molecular structure. We present results for a variety of different surface patterns in the supporting information (SI).

The results presented in Fig. 3 verify that $\delta\lambda_\text{phob}$ is effective as a local order parameter for surface hydrophobicity. This order parameter can thus be used to infer the effective hydration properties of an unknown aqueous surface. For the model silica surfaces the heterogeneity in surface chemistry is mirrored by the spatial dependence of $\delta\lambda_\text{phob}$, however, for more complex surfaces, the relationship between surface structure and $\delta\lambda_\text{phob}$ is not so straightforward. When this is the case, $\delta\lambda_\text{phob}$ can provide valuable physical insight into the relationship between surface heterogeneity and local hydration properties.

\begin{figure}[t!]
\centering
\includegraphics[width = 3.4 in]{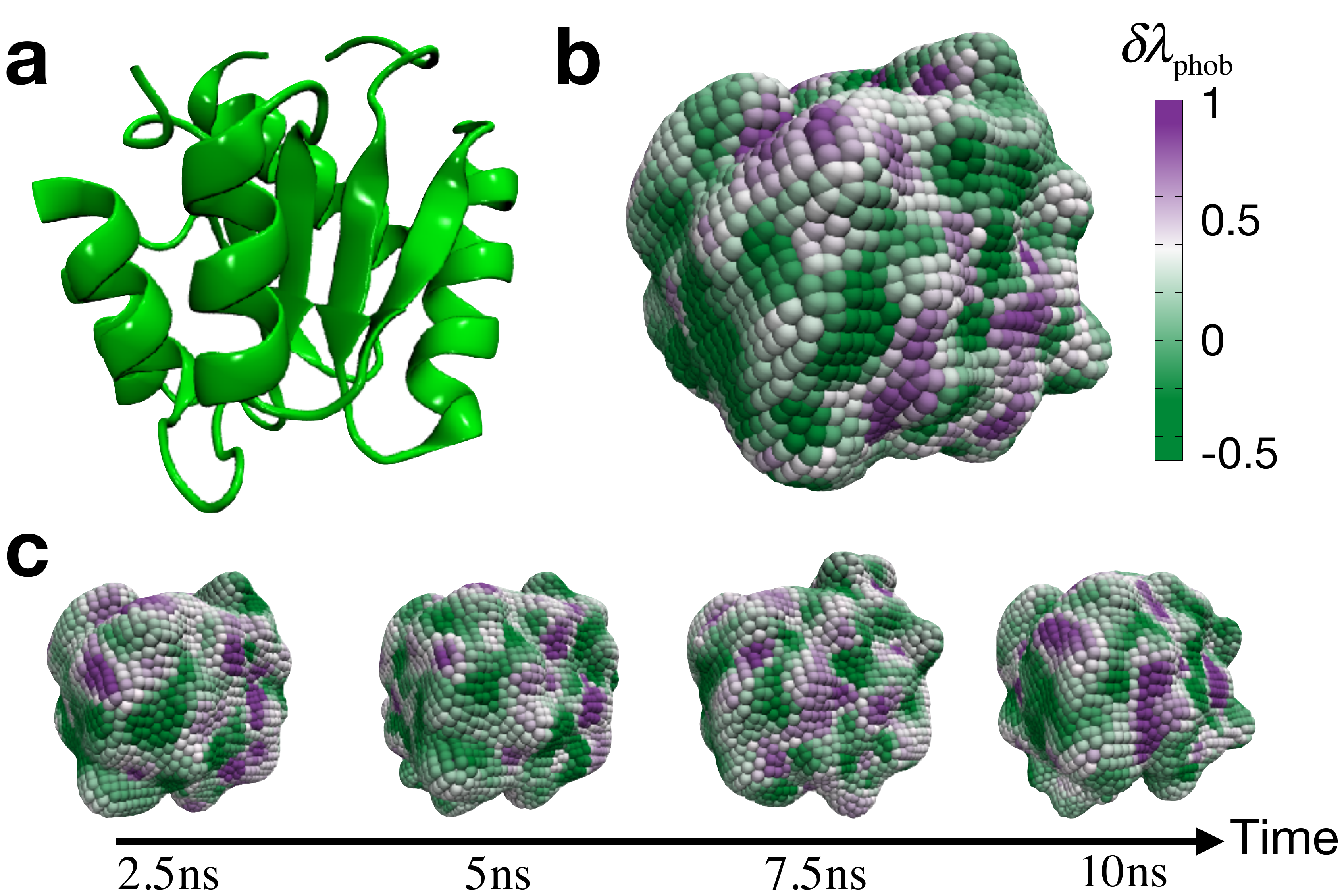}
\caption{(a) A simulation snapshot of the CheY protein. Water molecules are omitted for clarity. (b) A map of $\delta\lambda_\text{phob}$ computed for points along the surface of the protein using $\tau = 10 \text{ ps}$. Color scale is identical to that in Fig. 3b,d. (c) Spatial maps of $\delta\lambda_\text{phob}$ for a series of protein conformations spaced out along a 4 ns trajectory.}
\label{fig:4}
\end{figure}

The complex heterogeneous surface properties of hydrated proteins are reflected in the water's spatial dependence on interfacial molecular structure. Figure 4 illustrates that this spatial dependence can be revealed with $\delta\lambda_\text{phob}$. Specifically, Fig. 4 illustrates the value of $\delta\lambda_\text{phob}$ computed along the surface of the inactive CheY protein (PDB code: 1JBE) \cite{Simonovic:2001fp} using Eq. (\ref{eq:3}) and a value of $\tau=10 \text{ ps}$. This map of $\delta\lambda_\text{phob}$ indicates regions of the protein surface whose interactions with water result in hydrophobic (\emph{i.e.}, green shaded regions) or hydrophilic (\emph{i.e.}, purple shaded regions) interfacial molecular structure.

Details of these protein surface maps, such as the position and shapes of the hydrophilic domains, are sensitive to the conformational fluctuations of the protein. This is illustrated in Fig. 4c, which shows the map of $\delta\lambda_\text{phob}$ computed for the CheY protein for a consecutive sequence of conformations. The dynamics of interfacial structure can be further analyzed by considering the time dependence of $\delta\lambda_\text{phob}$ for individual surface residues. Figure 5 shows the result of such a calculation performed along a 10 ns trajectory of the CheY protein. We observe that the interfacial structure in some regions of the protein remains relatively static, as indicated by persistent green or purple bands in Fig. 5a, while other regions of the protein exhibit interfacial structure that fluctuates significantly in response to protein conformational dynamics. 

\begin{figure}[t!]
\centering
\includegraphics[width = 3.4 in]{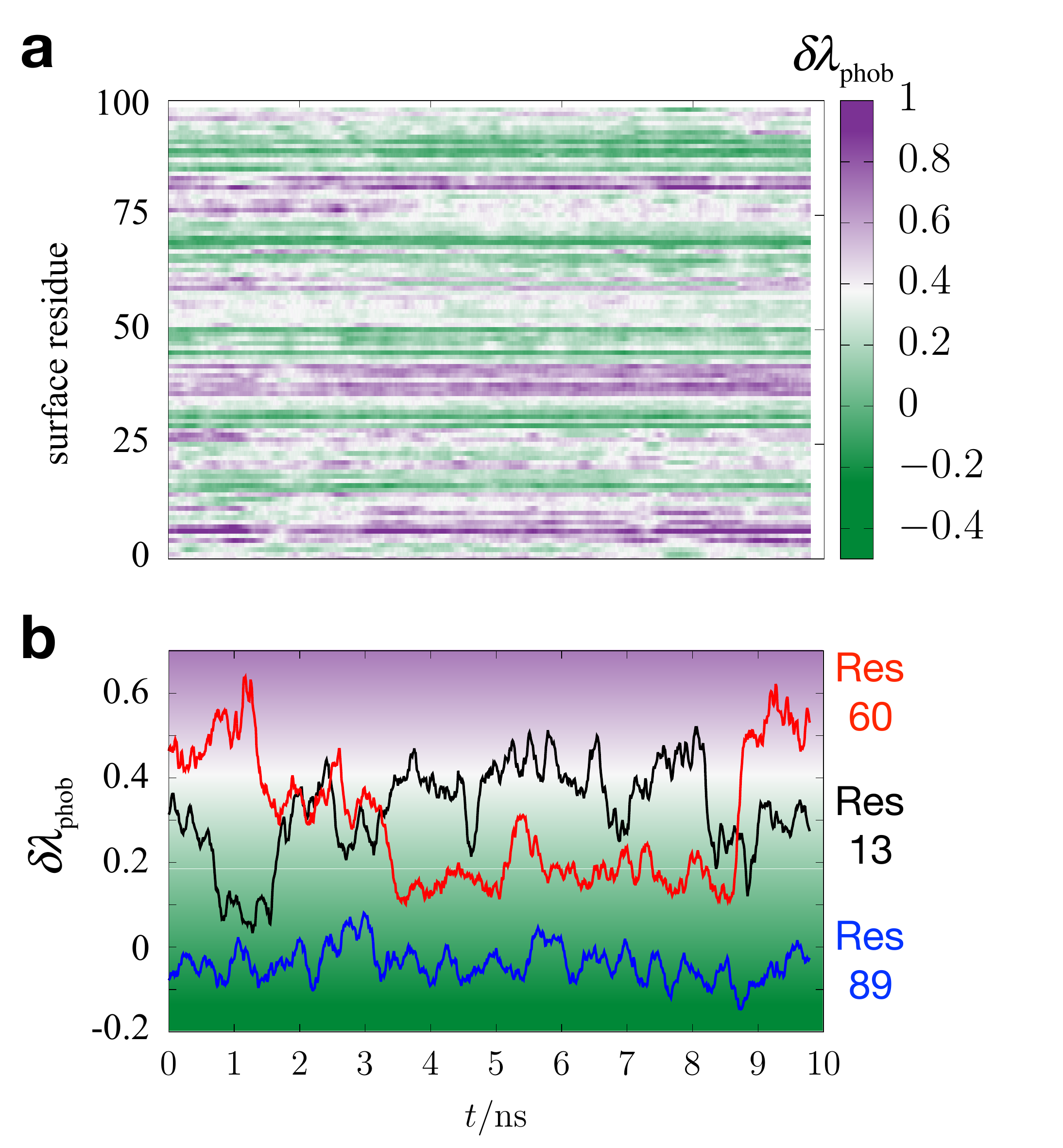}
\caption{(a) A time series plot of $\delta\lambda_\text{phob}$, indicated by shading, as computed for individual surface residues of the CheY protein. In this plot, each row corresponds to a unique surface residue. (b) A plot highlighting of the dynamics of $\delta\lambda_\text{phob}$ for three specific surface residues.}
\label{fig:5}
\end{figure}

The ability to map the hydration dynamics of fluctuating irregular solutes is a unique feature of this interfacial characterization method. Of course, this method is not limited to the applications presented above. For instance, the approach can be easily extended to report on additional hydration properties with the use of different reference systems, such as systematically charged surfaces or those with specific curvature, or by expanding the definition of $\vec{\kappa}$, for example to include dynamical information. Nor is the general approach limited to extended liquid water interfaces. The examples described here simply demonstrate the type of insight that can be derived from analyzing the orientational properties of interfacial water molecules.

\section{acknowledgement}
We acknowledge useful discussions with Professor John Weeks and Professor Gerhard Hummer. This work was supported by the National Science Foundation under CHE-1654415 and also partially (SS) by the Kwanjeong Educational Foundation in Korea.

\bibliography{q_bib.bib}

\begin{thebibliography}{25}%
\makeatletter
\providecommand \@ifxundefined [1]{%
 \@ifx{#1\undefined}
}%
\providecommand \@ifnum [1]{%
 \ifnum #1\expandafter \@firstoftwo
 \else \expandafter \@secondoftwo
 \fi
}%
\providecommand \@ifx [1]{%
 \ifx #1\expandafter \@firstoftwo
 \else \expandafter \@secondoftwo
 \fi
}%
\providecommand \natexlab [1]{#1}%
\providecommand \enquote  [1]{``#1''}%
\providecommand \bibnamefont  [1]{#1}%
\providecommand \bibfnamefont [1]{#1}%
\providecommand \citenamefont [1]{#1}%
\providecommand \href@noop [0]{\@secondoftwo}%
\providecommand \href [0]{\begingroup \@sanitize@url \@href}%
\providecommand \@href[1]{\@@startlink{#1}\@@href}%
\providecommand \@@href[1]{\endgroup#1\@@endlink}%
\providecommand \@sanitize@url [0]{\catcode `\\12\catcode `\$12\catcode
  `\&12\catcode `\#12\catcode `\^12\catcode `\_12\catcode `\%12\relax}%
\providecommand \@@startlink[1]{}%
\providecommand \@@endlink[0]{}%
\providecommand \url  [0]{\begingroup\@sanitize@url \@url }%
\providecommand \@url [1]{\endgroup\@href {#1}{\urlprefix }}%
\providecommand \urlprefix  [0]{URL }%
\providecommand \Eprint [0]{\href }%
\providecommand \doibase [0]{http://dx.doi.org/}%
\providecommand \selectlanguage [0]{\@gobble}%
\providecommand \bibinfo  [0]{\@secondoftwo}%
\providecommand \bibfield  [0]{\@secondoftwo}%
\providecommand \translation [1]{[#1]}%
\providecommand \BibitemOpen [0]{}%
\providecommand \bibitemStop [0]{}%
\providecommand \bibitemNoStop [0]{.\EOS\space}%
\providecommand \EOS [0]{\spacefactor3000\relax}%
\providecommand \BibitemShut  [1]{\csname bibitem#1\endcsname}%
\let\auto@bib@innerbib\@empty
\bibitem [{\citenamefont {Eisenthal}(1996)}]{Eisenthal:1996jh}%
  \BibitemOpen
  \bibfield  {author} {\bibinfo {author} {\bibfnamefont {K.~B.}\ \bibnamefont
  {Eisenthal}},\ }\href@noop {} {\bibfield  {journal} {\bibinfo  {journal}
  {Chemical reviews}\ }\textbf {\bibinfo {volume} {96}},\ \bibinfo {pages}
  {1343} (\bibinfo {year} {1996})}\BibitemShut {NoStop}%
\bibitem [{\citenamefont {Qu{\'e}r{\'e}}(2008)}]{Quere:2008cw}%
  \BibitemOpen
  \bibfield  {author} {\bibinfo {author} {\bibfnamefont {D.}~\bibnamefont
  {Qu{\'e}r{\'e}}},\ }\href@noop {} {\bibfield  {journal} {\bibinfo  {journal}
  {Annual Review of Materials Research}\ }\textbf {\bibinfo {volume} {38}},\
  \bibinfo {pages} {71} (\bibinfo {year} {2008})}\BibitemShut {NoStop}%
\bibitem [{\citenamefont {Verdaguer}\ \emph {et~al.}(2006)\citenamefont
  {Verdaguer}, \citenamefont {Sacha}, \citenamefont {Bluhm},\ and\
  \citenamefont {Salmeron}}]{Verdaguer:2006gl}%
  \BibitemOpen
  \bibfield  {author} {\bibinfo {author} {\bibfnamefont {A.}~\bibnamefont
  {Verdaguer}}, \bibinfo {author} {\bibfnamefont {G.~M.}\ \bibnamefont
  {Sacha}}, \bibinfo {author} {\bibfnamefont {H.}~\bibnamefont {Bluhm}}, \ and\
  \bibinfo {author} {\bibfnamefont {M.}~\bibnamefont {Salmeron}},\ }\href@noop
  {} {\bibfield  {journal} {\bibinfo  {journal} {Chemical reviews}\ }\textbf
  {\bibinfo {volume} {106}},\ \bibinfo {pages} {1478} (\bibinfo {year}
  {2006})}\BibitemShut {NoStop}%
\bibitem [{\citenamefont {Du}\ \emph {et~al.}(1993)\citenamefont {Du},
  \citenamefont {Superfine}, \citenamefont {Freysz},\ and\ \citenamefont
  {Shen}}]{Du:1993je}%
  \BibitemOpen
  \bibfield  {author} {\bibinfo {author} {\bibfnamefont {Q.}~\bibnamefont
  {Du}}, \bibinfo {author} {\bibfnamefont {R.}~\bibnamefont {Superfine}},
  \bibinfo {author} {\bibfnamefont {E.}~\bibnamefont {Freysz}}, \ and\ \bibinfo
  {author} {\bibfnamefont {Y.~R.}\ \bibnamefont {Shen}},\ }\href@noop {}
  {\bibfield  {journal} {\bibinfo  {journal} {Physical Review Letters}\
  }\textbf {\bibinfo {volume} {70}},\ \bibinfo {pages} {2313} (\bibinfo {year}
  {1993})}\BibitemShut {NoStop}%
\bibitem [{\citenamefont {Schleeger}\ \emph {et~al.}(2014)\citenamefont
  {Schleeger}, \citenamefont {Nagata},\ and\ \citenamefont
  {Bonn}}]{Schleeger:2014ek}%
  \BibitemOpen
  \bibfield  {author} {\bibinfo {author} {\bibfnamefont {M.}~\bibnamefont
  {Schleeger}}, \bibinfo {author} {\bibfnamefont {Y.}~\bibnamefont {Nagata}}, \
  and\ \bibinfo {author} {\bibfnamefont {M.}~\bibnamefont {Bonn}},\ }\href@noop
  {} {\bibfield  {journal} {\bibinfo  {journal} {The Journal of Physical
  Chemistry Letters}\ }\textbf {\bibinfo {volume} {5}},\ \bibinfo {pages}
  {3737} (\bibinfo {year} {2014})}\BibitemShut {NoStop}%
\bibitem [{\citenamefont {Singh}(2012)}]{Singh:2012kc}%
  \BibitemOpen
  \bibfield  {author} {\bibinfo {author} {\bibfnamefont {P.~C.}\ \bibnamefont
  {Singh}},\ }\href@noop {} {\bibfield  {journal} {\bibinfo  {journal} {The
  Journal of chemical physics}\ }\textbf {\bibinfo {volume} {137}},\ \bibinfo
  {pages} {094706} (\bibinfo {year} {2012})}\BibitemShut {NoStop}%
\bibitem [{\citenamefont {Baldelli}\ \emph {et~al.}(2002)\citenamefont
  {Baldelli}, \citenamefont {Schnitzer},\ and\ \citenamefont
  {Simonelli}}]{Baldelli:2002be}%
  \BibitemOpen
  \bibfield  {author} {\bibinfo {author} {\bibfnamefont {S.}~\bibnamefont
  {Baldelli}}, \bibinfo {author} {\bibfnamefont {C.}~\bibnamefont {Schnitzer}},
  \ and\ \bibinfo {author} {\bibfnamefont {D.}~\bibnamefont {Simonelli}},\
  }\href@noop {} {\bibfield  {journal} {\bibinfo  {journal} {The Journal of
  Physical Chemistry B}\ }\textbf {\bibinfo {volume} {106}},\ \bibinfo {pages}
  {5313} (\bibinfo {year} {2002})}\BibitemShut {NoStop}%
\bibitem [{\citenamefont {Gragson}\ \emph {et~al.}(1997)\citenamefont
  {Gragson}, \citenamefont {McCarty},\ and\ \citenamefont
  {Richmond}}]{Gragson:1997dj}%
  \BibitemOpen
  \bibfield  {author} {\bibinfo {author} {\bibfnamefont {D.~E.}\ \bibnamefont
  {Gragson}}, \bibinfo {author} {\bibfnamefont {B.~M.}\ \bibnamefont
  {McCarty}}, \ and\ \bibinfo {author} {\bibfnamefont {G.~L.}\ \bibnamefont
  {Richmond}},\ }\href@noop {} {\bibfield  {journal} {\bibinfo  {journal}
  {Journal of the American Chemical Society}\ }\textbf {\bibinfo {volume}
  {119}},\ \bibinfo {pages} {6144} (\bibinfo {year} {1997})}\BibitemShut
  {NoStop}%
\bibitem [{\citenamefont {Conti~Nibali}\ and\ \citenamefont
  {Havenith}(2014)}]{ContiNibali:2014gt}%
  \BibitemOpen
  \bibfield  {author} {\bibinfo {author} {\bibfnamefont {V.}~\bibnamefont
  {Conti~Nibali}}\ and\ \bibinfo {author} {\bibfnamefont {M.}~\bibnamefont
  {Havenith}},\ }\href@noop {} {\bibfield  {journal} {\bibinfo  {journal}
  {Journal of the American Chemical Society}\ }\textbf {\bibinfo {volume}
  {136}},\ \bibinfo {pages} {12800} (\bibinfo {year} {2014})}\BibitemShut
  {NoStop}%
\bibitem [{\citenamefont {Ebbinghaus}\ \emph {et~al.}(2007)\citenamefont
  {Ebbinghaus}, \citenamefont {Kim}, \citenamefont {Heyden}, \citenamefont
  {Yu}, \citenamefont {Heugen}, \citenamefont {Gruebele}, \citenamefont
  {Leitner},\ and\ \citenamefont {Havenith}}]{Ebbinghaus:2007gl}%
  \BibitemOpen
  \bibfield  {author} {\bibinfo {author} {\bibfnamefont {S.}~\bibnamefont
  {Ebbinghaus}}, \bibinfo {author} {\bibfnamefont {S.~J.}\ \bibnamefont {Kim}},
  \bibinfo {author} {\bibfnamefont {M.}~\bibnamefont {Heyden}}, \bibinfo
  {author} {\bibfnamefont {X.}~\bibnamefont {Yu}}, \bibinfo {author}
  {\bibfnamefont {U.}~\bibnamefont {Heugen}}, \bibinfo {author} {\bibfnamefont
  {M.}~\bibnamefont {Gruebele}}, \bibinfo {author} {\bibfnamefont {D.~M.}\
  \bibnamefont {Leitner}}, \ and\ \bibinfo {author} {\bibfnamefont
  {M.}~\bibnamefont {Havenith}},\ }\href@noop {} {\bibfield  {journal}
  {\bibinfo  {journal} {Proceedings of the National Academy of Sciences}\
  }\textbf {\bibinfo {volume} {104}},\ \bibinfo {pages} {20749} (\bibinfo
  {year} {2007})}\BibitemShut {NoStop}%
\bibitem [{\citenamefont {Armstrong}\ and\ \citenamefont
  {Han}(2009)}]{Armstrong:2009et}%
  \BibitemOpen
  \bibfield  {author} {\bibinfo {author} {\bibfnamefont {B.~D.}\ \bibnamefont
  {Armstrong}}\ and\ \bibinfo {author} {\bibfnamefont {S.}~\bibnamefont
  {Han}},\ }\href@noop {} {\bibfield  {journal} {\bibinfo  {journal} {Journal
  of the American Chemical Society}\ }\textbf {\bibinfo {volume} {131}},\
  \bibinfo {pages} {11270} (\bibinfo {year} {2009})}\BibitemShut {NoStop}%
\bibitem [{\citenamefont {Otting}\ and\ \citenamefont
  {Wuethrich}(1989)}]{Otting:1989jv}%
  \BibitemOpen
  \bibfield  {author} {\bibinfo {author} {\bibfnamefont {G.}~\bibnamefont
  {Otting}}\ and\ \bibinfo {author} {\bibfnamefont {K.}~\bibnamefont
  {Wuethrich}},\ }\href@noop {} {\bibfield  {journal} {\bibinfo  {journal}
  {Journal of the American Chemical Society}\ }\textbf {\bibinfo {volume}
  {111}},\ \bibinfo {pages} {1871} (\bibinfo {year} {1989})}\BibitemShut
  {NoStop}%
\bibitem [{\citenamefont {Otting}\ \emph {et~al.}(1991)\citenamefont {Otting},
  \citenamefont {Liepinsh},\ and\ \citenamefont
  {W{\"u}thrich}}]{Otting:1991ue}%
  \BibitemOpen
  \bibfield  {author} {\bibinfo {author} {\bibfnamefont {G.}~\bibnamefont
  {Otting}}, \bibinfo {author} {\bibfnamefont {E.}~\bibnamefont {Liepinsh}}, \
  and\ \bibinfo {author} {\bibfnamefont {K.}~\bibnamefont {W{\"u}thrich}},\
  }\href@noop {} {\bibfield  {journal} {\bibinfo  {journal} {Science (New York,
  N.Y.)}\ }\textbf {\bibinfo {volume} {254}},\ \bibinfo {pages} {974} (\bibinfo
  {year} {1991})}\BibitemShut {NoStop}%
\bibitem [{\citenamefont {Svergun}\ \emph {et~al.}(1998)\citenamefont
  {Svergun}, \citenamefont {Richard}, \citenamefont {Koch}, \citenamefont
  {Sayers}, \citenamefont {Kuprin},\ and\ \citenamefont
  {Zaccai}}]{Svergun:1998ej}%
  \BibitemOpen
  \bibfield  {author} {\bibinfo {author} {\bibfnamefont {D.~I.}\ \bibnamefont
  {Svergun}}, \bibinfo {author} {\bibfnamefont {S.}~\bibnamefont {Richard}},
  \bibinfo {author} {\bibfnamefont {M.~H.~J.}\ \bibnamefont {Koch}}, \bibinfo
  {author} {\bibfnamefont {Z.}~\bibnamefont {Sayers}}, \bibinfo {author}
  {\bibfnamefont {S.}~\bibnamefont {Kuprin}}, \ and\ \bibinfo {author}
  {\bibfnamefont {G.}~\bibnamefont {Zaccai}},\ }\href@noop {} {\bibfield
  {journal} {\bibinfo  {journal} {Proceedings of the National Academy of
  Sciences}\ }\textbf {\bibinfo {volume} {95}},\ \bibinfo {pages} {2267}
  (\bibinfo {year} {1998})}\BibitemShut {NoStop}%
\bibitem [{\citenamefont {Bruni}\ \emph {et~al.}(1998)\citenamefont {Bruni},
  \citenamefont {Ricci},\ and\ \citenamefont {Soper}}]{Bruni:1998gy}%
  \BibitemOpen
  \bibfield  {author} {\bibinfo {author} {\bibfnamefont {F.}~\bibnamefont
  {Bruni}}, \bibinfo {author} {\bibfnamefont {M.~A.}\ \bibnamefont {Ricci}}, \
  and\ \bibinfo {author} {\bibfnamefont {A.~K.}\ \bibnamefont {Soper}},\
  }\href@noop {} {\bibfield  {journal} {\bibinfo  {journal} {The Journal of
  chemical physics}\ }\textbf {\bibinfo {volume} {109}},\ \bibinfo {pages}
  {1478} (\bibinfo {year} {1998})}\BibitemShut {NoStop}%
\bibitem [{\citenamefont {Luo}\ \emph {et~al.}(2006)\citenamefont {Luo},
  \citenamefont {Malkova}, \citenamefont {Yoon}, \citenamefont {Schultz},
  \citenamefont {Lin}, \citenamefont {Meron}, \citenamefont {Benjamin},
  \citenamefont {Vanysek},\ and\ \citenamefont {Schlossman}}]{Luo:2006gu}%
  \BibitemOpen
  \bibfield  {author} {\bibinfo {author} {\bibfnamefont {G.}~\bibnamefont
  {Luo}}, \bibinfo {author} {\bibfnamefont {S.}~\bibnamefont {Malkova}},
  \bibinfo {author} {\bibfnamefont {J.}~\bibnamefont {Yoon}}, \bibinfo {author}
  {\bibfnamefont {D.~G.}\ \bibnamefont {Schultz}}, \bibinfo {author}
  {\bibfnamefont {B.}~\bibnamefont {Lin}}, \bibinfo {author} {\bibfnamefont
  {M.}~\bibnamefont {Meron}}, \bibinfo {author} {\bibfnamefont
  {I.}~\bibnamefont {Benjamin}}, \bibinfo {author} {\bibfnamefont
  {P.}~\bibnamefont {Vanysek}}, \ and\ \bibinfo {author} {\bibfnamefont
  {M.~L.}\ \bibnamefont {Schlossman}},\ }\href@noop {} {\bibfield  {journal}
  {\bibinfo  {journal} {Science (New York, N.Y.)}\ }\textbf {\bibinfo {volume}
  {311}},\ \bibinfo {pages} {216} (\bibinfo {year} {2006})}\BibitemShut
  {NoStop}%
\bibitem [{\citenamefont {Willard}\ and\ \citenamefont
  {Chandler}(2010)}]{Willard:2010da}%
  \BibitemOpen
  \bibfield  {author} {\bibinfo {author} {\bibfnamefont {A.~P.}\ \bibnamefont
  {Willard}}\ and\ \bibinfo {author} {\bibfnamefont {D.}~\bibnamefont
  {Chandler}},\ }\href@noop {} {\bibfield  {journal} {\bibinfo  {journal} {The
  Journal of Physical Chemistry B}\ }\textbf {\bibinfo {volume} {114}},\
  \bibinfo {pages} {1954} (\bibinfo {year} {2010})}\BibitemShut {NoStop}%
\bibitem [{\citenamefont {Kyte}\ and\ \citenamefont
  {Doolittle}(1982)}]{Kyte:1982fr}%
  \BibitemOpen
  \bibfield  {author} {\bibinfo {author} {\bibfnamefont {J.}~\bibnamefont
  {Kyte}}\ and\ \bibinfo {author} {\bibfnamefont {R.~F.}\ \bibnamefont
  {Doolittle}},\ }\href@noop {} {\bibfield  {journal} {\bibinfo  {journal}
  {Journal of molecular biology}\ }\textbf {\bibinfo {volume} {157}},\ \bibinfo
  {pages} {105} (\bibinfo {year} {1982})}\BibitemShut {NoStop}%
\bibitem [{\citenamefont {Kapcha}\ and\ \citenamefont
  {Rossky}(2014)}]{Kapcha:2014fg}%
  \BibitemOpen
  \bibfield  {author} {\bibinfo {author} {\bibfnamefont {L.~H.}\ \bibnamefont
  {Kapcha}}\ and\ \bibinfo {author} {\bibfnamefont {P.~J.}\ \bibnamefont
  {Rossky}},\ }\href@noop {} {\bibfield  {journal} {\bibinfo  {journal}
  {Journal of molecular biology}\ }\textbf {\bibinfo {volume} {426}},\ \bibinfo
  {pages} {484} (\bibinfo {year} {2014})}\BibitemShut {NoStop}%
\bibitem [{\citenamefont {Limmer}\ and\ \citenamefont
  {Willard}(2015)}]{Limmer:2015bq}%
  \BibitemOpen
  \bibfield  {author} {\bibinfo {author} {\bibfnamefont {D.~T.}\ \bibnamefont
  {Limmer}}\ and\ \bibinfo {author} {\bibfnamefont {A.~P.}\ \bibnamefont
  {Willard}},\ }\href@noop {} {\bibfield  {journal} {\bibinfo  {journal}
  {Chemical Physics Letters}\ }\textbf {\bibinfo {volume} {620}},\ \bibinfo
  {pages} {144} (\bibinfo {year} {2015})}\BibitemShut {NoStop}%
\bibitem [{\citenamefont {Patel}\ \emph {et~al.}(2011)\citenamefont {Patel},
  \citenamefont {Varilly}, \citenamefont {Chandler},\ and\ \citenamefont
  {Garde}}]{Patel:2011dz}%
  \BibitemOpen
  \bibfield  {author} {\bibinfo {author} {\bibfnamefont {A.~J.}\ \bibnamefont
  {Patel}}, \bibinfo {author} {\bibfnamefont {P.}~\bibnamefont {Varilly}},
  \bibinfo {author} {\bibfnamefont {D.}~\bibnamefont {Chandler}}, \ and\
  \bibinfo {author} {\bibfnamefont {S.}~\bibnamefont {Garde}},\ }\href@noop {}
  {\bibfield  {journal} {\bibinfo  {journal} {Journal of statistical physics}\
  }\textbf {\bibinfo {volume} {145}},\ \bibinfo {pages} {265} (\bibinfo {year}
  {2011})}\BibitemShut {NoStop}%
\bibitem [{\citenamefont {Young}\ \emph {et~al.}(2007)\citenamefont {Young},
  \citenamefont {Abel}, \citenamefont {Kim}, \citenamefont {Berne},\ and\
  \citenamefont {Friesner}}]{Young:2007cx}%
  \BibitemOpen
  \bibfield  {author} {\bibinfo {author} {\bibfnamefont {T.}~\bibnamefont
  {Young}}, \bibinfo {author} {\bibfnamefont {R.}~\bibnamefont {Abel}},
  \bibinfo {author} {\bibfnamefont {B.}~\bibnamefont {Kim}}, \bibinfo {author}
  {\bibfnamefont {B.~J.}\ \bibnamefont {Berne}}, \ and\ \bibinfo {author}
  {\bibfnamefont {R.~A.}\ \bibnamefont {Friesner}},\ }\href@noop {} {\bibfield
  {journal} {\bibinfo  {journal} {Proceedings of the National Academy of
  Sciences}\ }\textbf {\bibinfo {volume} {104}},\ \bibinfo {pages} {808}
  (\bibinfo {year} {2007})}\BibitemShut {NoStop}%
\bibitem [{\citenamefont {Remsing}\ and\ \citenamefont
  {Weeks}(2015)}]{Remsing:2015dh}%
  \BibitemOpen
  \bibfield  {author} {\bibinfo {author} {\bibfnamefont {R.~C.}\ \bibnamefont
  {Remsing}}\ and\ \bibinfo {author} {\bibfnamefont {J.~D.}\ \bibnamefont
  {Weeks}},\ }\href@noop {} {\bibfield  {journal} {\bibinfo  {journal} {The
  Journal of Physical Chemistry B}\ }\textbf {\bibinfo {volume} {119}},\
  \bibinfo {pages} {9268} (\bibinfo {year} {2015})}\BibitemShut {NoStop}%
\bibitem [{\citenamefont {Giovambattista}\ \emph {et~al.}(2007)\citenamefont
  {Giovambattista}, \citenamefont {Debenedetti},\ and\ \citenamefont
  {Rossky}}]{Giovambattista:2007cj}%
  \BibitemOpen
  \bibfield  {author} {\bibinfo {author} {\bibfnamefont {N.}~\bibnamefont
  {Giovambattista}}, \bibinfo {author} {\bibfnamefont {P.~G.}\ \bibnamefont
  {Debenedetti}}, \ and\ \bibinfo {author} {\bibfnamefont {P.~J.}\ \bibnamefont
  {Rossky}},\ }\href@noop {} {\bibfield  {journal} {\bibinfo  {journal} {The
  Journal of Physical Chemistry C}\ }\textbf {\bibinfo {volume} {111}},\
  \bibinfo {pages} {1323} (\bibinfo {year} {2007})}\BibitemShut {NoStop}%
\bibitem [{\citenamefont {Simonovic}\ and\ \citenamefont
  {Volz}(2001)}]{Simonovic:2001fp}%
  \BibitemOpen
  \bibfield  {author} {\bibinfo {author} {\bibfnamefont {M.}~\bibnamefont
  {Simonovic}}\ and\ \bibinfo {author} {\bibfnamefont {K.}~\bibnamefont
  {Volz}},\ }\href@noop {} {\bibfield  {journal} {\bibinfo  {journal} {The
  Journal of biological chemistry}\ }\textbf {\bibinfo {volume} {276}},\
  \bibinfo {pages} {28637} (\bibinfo {year} {2001})}\BibitemShut {NoStop}%
\end{thebibliography}%

\clearpage

\begin{figure}[t!]
\centering
\includegraphics[width = 3.4 in]{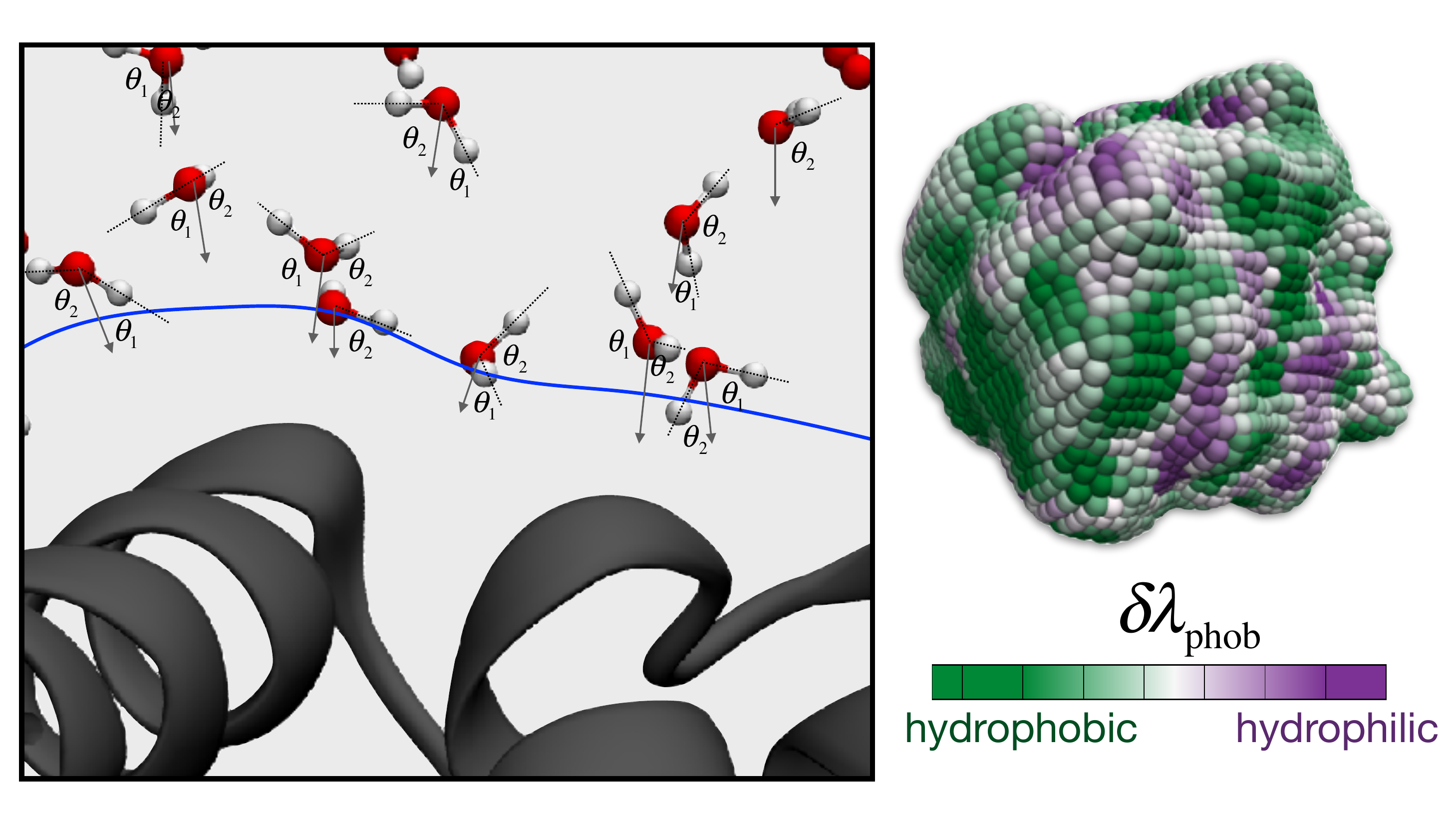}
\caption{Table of Contents graphic.}
\label{fig:TOC}
\end{figure}

\end{document}